\documentclass[prl,aps,superscriptaddress,
floatfix,tightenlines,twocolumn]{revtex4}

\usepackage{amsmath}
\usepackage{amssymb}
\usepackage{amsfonts}
\usepackage{subfigure}
\usepackage[dvips]{color}
\usepackage[colorlinks,bookmarks=false,citecolor=blue,linkcolor=red,urlcolor=blue]{hyperref}
\usepackage{bm}
\usepackage{amsthm}
\usepackage{graphicx}
\usepackage{multirow}
\usepackage{mathrsfs}
\usepackage{times}

\theoremstyle{plain}

\theoremstyle{definition}

\theoremstyle{remark}

\begin{document}

\title{Experimental valley qubit state tomography and coherence induced uncertainty relations in monolayer WSe$_2$}
\author{Yu-Ran Zhang}
\affiliation{Institute of Physics, Chinese Academy of Sciences, Beijing 100190, China}
\affiliation{School of Physical Sciences, University of Chinese Academy of Sciences, Beijing 100190, China}
\author{{Jing Wang}}
\affiliation{Institute of Physics, Chinese Academy of Sciences, Beijing 100190, China}
\affiliation{School of Physical Sciences, University of Chinese Academy of Sciences, Beijing 100190, China}
\author{{Chuanrui Zhu}}
\affiliation{Institute of Physics, Chinese Academy of Sciences, Beijing 100190, China}
\affiliation{School of Physical Sciences, University of Chinese Academy of Sciences, Beijing 100190, China}
\author{Heng Fan}
\email{hfan@iphy.ac.cn}
\affiliation{Institute of Physics, Chinese Academy of Sciences, Beijing 100190, China}
\affiliation{School of Physical Sciences, University of Chinese Academy of Sciences, Beijing 100190, China}
\affiliation{Collaborative Innovation Center of Quantum Matter, Beijing 100190, China}
\author{Baoli Liu}
\email{blliu@iphy.ac.cn}
\affiliation{Institute of Physics, Chinese Academy of Sciences, Beijing 100190, China}

\maketitle

{\bf


Velleytronics as a new electronic conception is an emerging exciting research field with wide potential applications,
which is attracting great research interests for their extraordinary
properties \cite{Novoselov2005,Rycerz2007,Shkolnikov2002,Gunawan2006,Schaibley2016}.
The localized electronic spins by optical generation of valley
polarization \cite{Cao2012,Mak2012,Zeng2012,Jones2013} with spin-like quantum numbers
\cite{Xiao2012,Zhu2012,Bishop2007} are promising candidates for implementing quantum-information
processing in solids. It is expected that a single qubit preparation can be realized optically by using combination of left- and
right-circularly polarized lights \cite{Mak2012,Zeng2012,Cao2012}. Significantly in a series of experiments, this has already been well achieved by linearly polarized laser
 representing equal weights of left- and right-circular components resulting in formation of a {valley} exciton
\cite{Chernikov2014,Ugeda2014,He2014,Ye2014}
 as a specific
pseudo-spin qubit with equal amplitudes for spin up and spin down.
Further
researches on the control of valley pseudospin using longitudinal magnetic field \cite{Wang2016,Schmidt2016} and optical Stark effect \cite{Ye2016} have been reported.
However, a general qubit preparation has not yet
been demonstrated. Moreover as a platform for quantum information processing, the precise readout of a qubit state is necessary, for which the state tomography is a standard method in obtaining all information of a qubit state density matrix.
}

Here, we will lay all necessity foundations in quantum-information processing for the valley pseudospin as a qubit.
Our main results are as follows: we show that an arbitrary qubit preparation can be implemented by using
the specific elliptical light excitation with combination of the corresponding left- and right-circularly polarized lights. Then we show that the valley qubit readout of the valley pseudospin
can be achieved precisely by the standard state tomography by measuring
both linear and circular polarization degrees and intensities of the polarization-resolved photoluminescence (PL).
Thus the density matrix of the valley qubit is presented quantitatively.
To confirm the coherence of the valley qubit, we show that the valley pseudospin qubit can
demonstrate the Heisenberg uncertainty principle which the unique characteristic of
quantum mechanics, including two different forms: entropic uncertainty relations and
Heisenberg uncertainty relations.
Our results pave the way for the valley pseudospin as a qubit which is the
fundamental element acting as the carrier of quantum information.

\begin{figure}[t]
\includegraphics[width=0.48\textwidth]{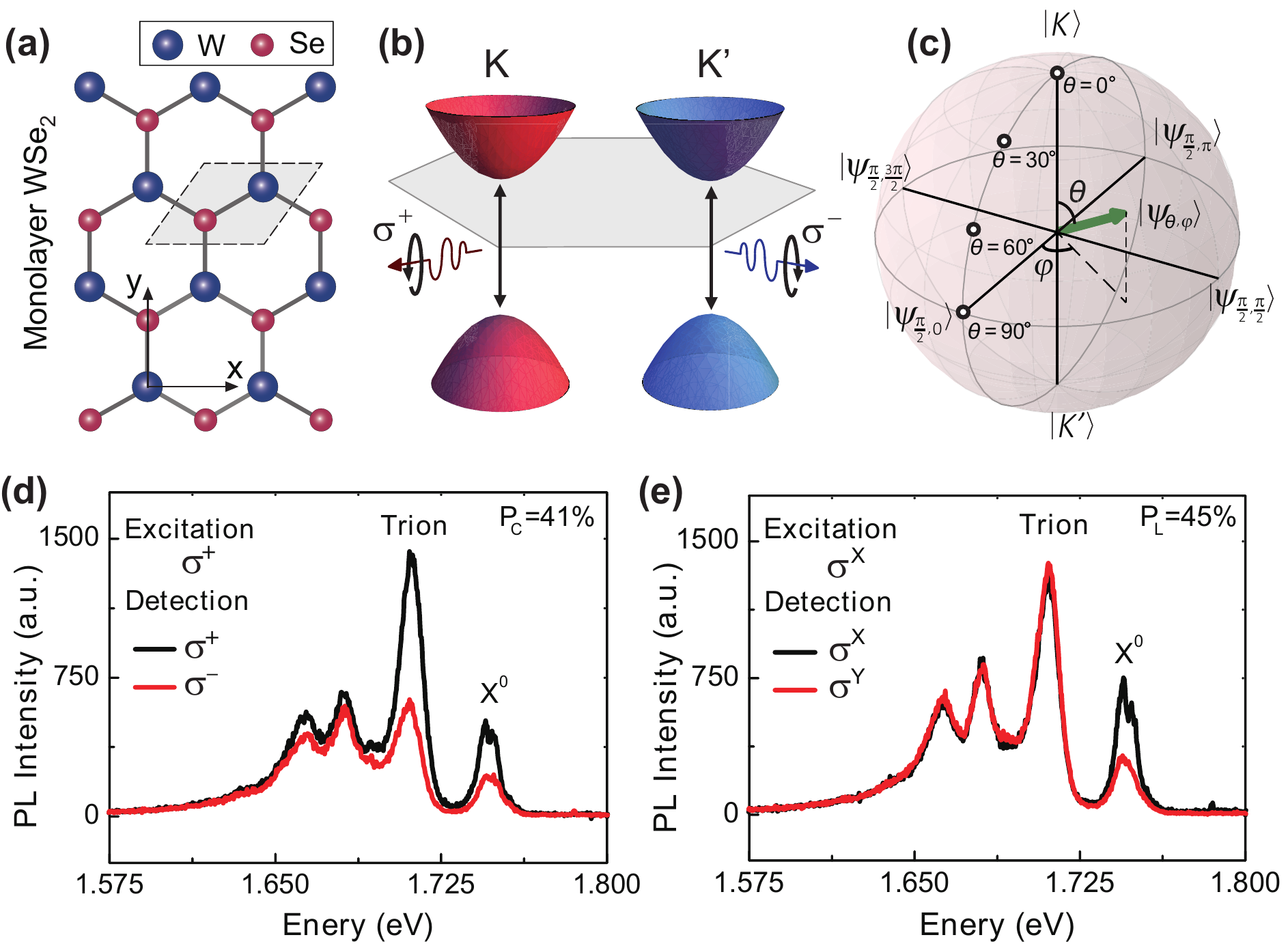}\\
\caption{\textbf{Valley-orbit coupled neutral exciton $X^0$ as a qubit.}
(a) The 2D hexagonal crystal structure of a monolayer (ML) transition metal dichalcogenide (TMD)
composed of W atoms (blue) and Se atoms (red).
(b) Valley optical selection rules for neutral exciton $X^0$ in ML TMD. In $K$ and $K'$ valleys, the neutral excitons $X^0$
recombine to emit $\sigma^+$ and $\sigma^-$ circularly polarized photons, respectively.
(c) Bloch sphere representation of the optical manipulation of  neutral excitonic pseudo-spin states.
(d) Photoluminescence (PL) spectra of ML WSe${}_2$ at 4.7~K under $\sigma^+$ circularly polarized excitation.
(e) PL spectra of ML WSe${}_2$ at 4.7~K under $\sigma^X$ linearly polarized excitation.}\label{fig:1}
\end{figure}

\textbf{Valley pseudospin as a qubit and its detection by photoluminescence.}
Tungsten diselenide (WSe${}_2$),  a two-dimensional ML of TMDs (MX$_2$), have emerged as an
exciting platform for opto-electronics and quantum information processing, see Fig.~\ref{fig:1}(a). In
particular, a pair of degenerate energy extremals are present at the $K$ and $K'$ valleys in the momentum
space of hexagonal MX$_2$ MLs. These valleys represented by a binary pseudospin behave like a
spin-1/2 system where the
neutral exciton  in the $K$ valley can be labelled as valley-pseudospin up, and
the
neutral exciton  in the $K'$ valley can be labelled as valley-pseudospin down \cite{Schaibley2016}. The quantum states of two valleys and their coherent superposition
constitute the valley pseudospin, promising the valley degree of freedom to realize a qubit. As optically driven spintronics, the generation and detection
of valley pseudospin depend on the facts that the left-circularly ($\sigma^{+}$) or right-circularly ($\sigma^{-}$)
polarized light selectively excites an electron-hole pair at the $K$ or $K'$ valley.
In turn, the resulting valley-polarized
exciton exclusively couples to left-circularly or right-circularly polarized light and
can be detected by polarization-resolved PL, see Fig.~\ref{fig:1}(b).

In our experiment, the  ML WSe${}_2$ flakes are prepared by mechanical exfoliation of bulk
WSe${}_2$ crystal (2D semiconductors USA) on SiO${}_{2}$/Si substrate. The samples are mounted in a
temperature-controlled (4.7~K $\sim$ 300~K) He continuous flow cryostat (Oxford MicrostatHiResII). The
measurements are carried out in confocal microscope with a super-long working distance (22~mm) objective
(Nikon CFI60-2, 50$\times$). The typical spot size of focused laser is around $\sim$2~$\mu$m and is smaller than
the size of ML WSe${}_2$. A 671~nm continuous wave (cw)
solid-state laser was used as an excitation light source cleaned by a 671~nm laser line filter.
The valley pseudospin polarization is realized by excitonic robust
bound states consisting of electrons and holes localized in the $K$ and $K'$ valleys, and it can
be optically manipulated and detected. The target valley states are initialized by elliptically-polarized
excitation laser light passing through a Soleil Babinet Compensator.

\begin{figure}[t]
 \centering
\includegraphics[width=0.43\textwidth]{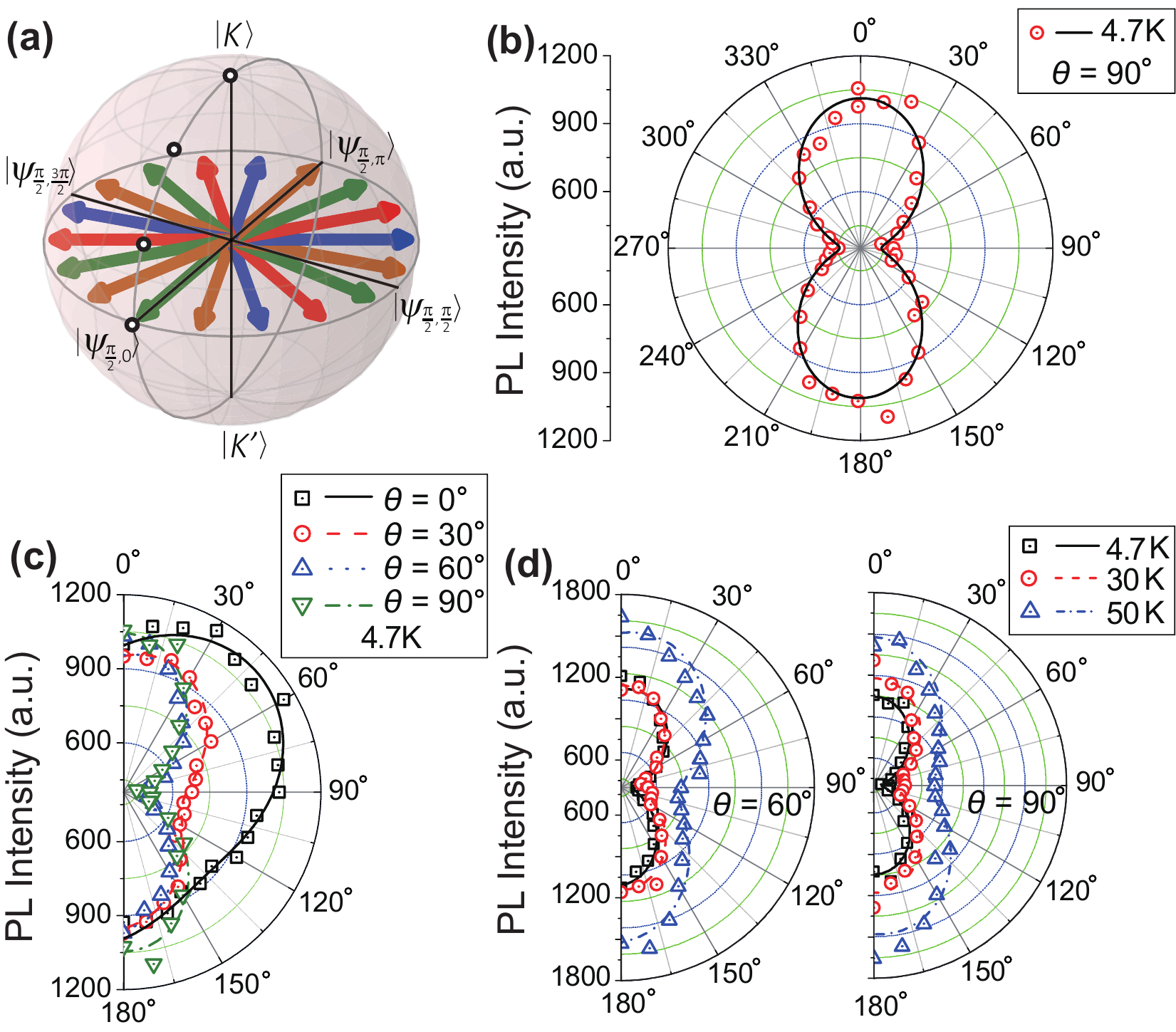}\\
\caption{\textbf{Signature of valley quantum coherence.}
(a)  To detect the excitonic valley quantum coherence, a series of equatorial projective measurements
are applied with projectors $\hat{\Pi}_\alpha\equiv|\psi_{\frac{\pi}{2},\alpha}\rangle\langle\psi_{\frac{\pi}{2},\alpha}|$
and with detection angles $\alpha$ from $0^\circ$ to $360^\circ$.
(b) PL intensity as a function of detection angles $\alpha$ when the neutral exciton state
$|\psi_{\frac{\pi}{2},0}\rangle$ is excited by $\sigma^X$ linearly polarized laser at 4.7~K.
(c) PL intensity as a function of detection angles $\alpha$ at 4.7~K for different neutral exciton states
$|\psi_{\theta,0}\rangle$ by linearly polarized laser with $\theta=0^\circ,30^\circ,60^\circ,90^\circ$, respectively.
(d) At different temperatures 4.7~K, 30~K and 50~K, PL intensity as a function of detection angles $\alpha$
where the left diagram is for a neutral exciton state $|\psi_{\frac{\pi}{2},0}\rangle$ and the right one  is for
a neutral exciton state $|\psi_{\frac{\pi}{2},0}\rangle$.
}\label{fig:2}
\end{figure}

The circular basis consists of $|K\rangle$ and $|K'\rangle$ valley states
which are at two poles in Bloch sphere representation, see Fig~\ref{fig:1}(c).
In the frame of Bloch sphere, an arbitrary pure valley state is written as
\begin{eqnarray}
|\psi_{\theta,\varphi}\rangle=\cos\frac{\theta}{2}|K\rangle+\sin\frac{\theta}{2}e^{i\varphi}|K'\rangle.\label{eq:1}
\end{eqnarray}
In order to prepare this state, we need to excite the system by the incident elliptically polarized light
{superposed by left- and right-circularly polarized lights with different weights and relative phase, in which
the weight and phase correspond
to $\theta$
 and $\varphi$, respectively.} The linearly polarized basis consists of two arbitrary orthogonal states located in the equator of Bloch sphere represented by $\sigma^{X,Y}$ corresponding to valley states
 $|\psi_{\frac{\pi}{2},\frac{\pi}{2}\mp\frac{\pi}{2}}\rangle=(|K\rangle\pm|K'\rangle)/\sqrt{2}$, respectively.

\textbf{Normalized neutral exciton measurement and valley state tomography.}
For the measurements of prepared valley states in the
basis of $|K\rangle$ and $|K'\rangle$ vectors, we detect the left- and right-circularly
circularly polarized PL intensities of neutral exciton $X^0$ and obtain the circular polarization
$\eta_{\textrm{C}}$ by measuring the circularly polarized components of the emitted PL intensity $I(\sigma^\pm)$ as \cite{Cao2012, Schaibley2016}
$\eta_\textrm{C}=[{I(\sigma^+)-I(\sigma^-)}]/[{I(\sigma^+)+I(\sigma^-)}]$.
Similarly, we can detect the valley linear polarization
$\eta_{\textrm{L}}$ by measuring the linearly polarized components of the emitted PL intensity $I(\sigma^{X,Y})$ as
$\eta_\textrm{L}=[{I(\sigma^X)-I(\sigma^Y)}]/[{I(\sigma^X)+I(\sigma^Y)}]$.
Figures~\ref{fig:1}(d,e) show the typical PL spectra of ML WSe${}_2$ at 4.7~K
under $\sigma^{+}$ circular and $\sigma^{X}$ linear polarization excitation, respectively. Both spectra show the higher
polarization of neutral exciton $X^0$.
To initialize any pure valley state (\ref{eq:1}),  degrees of circular polarization and
linear polarization of excitation light, are ideally $\eta_\textrm{C}=\cos\theta$ and $\eta_\textrm{L}=\sin\theta$.
 In this paper, we have prepared four valley states
with $\varphi=0$ and $\theta=0^\circ,30^\circ,60^\circ,90^\circ$, respectively.
The normalized population of polar projective measurement can be obtained
by using experimental degrees of both circular and polarization of PL, see Supplementary Information for details.

\begin{figure*}[t]
\centering
\includegraphics[width=0.96\textwidth]{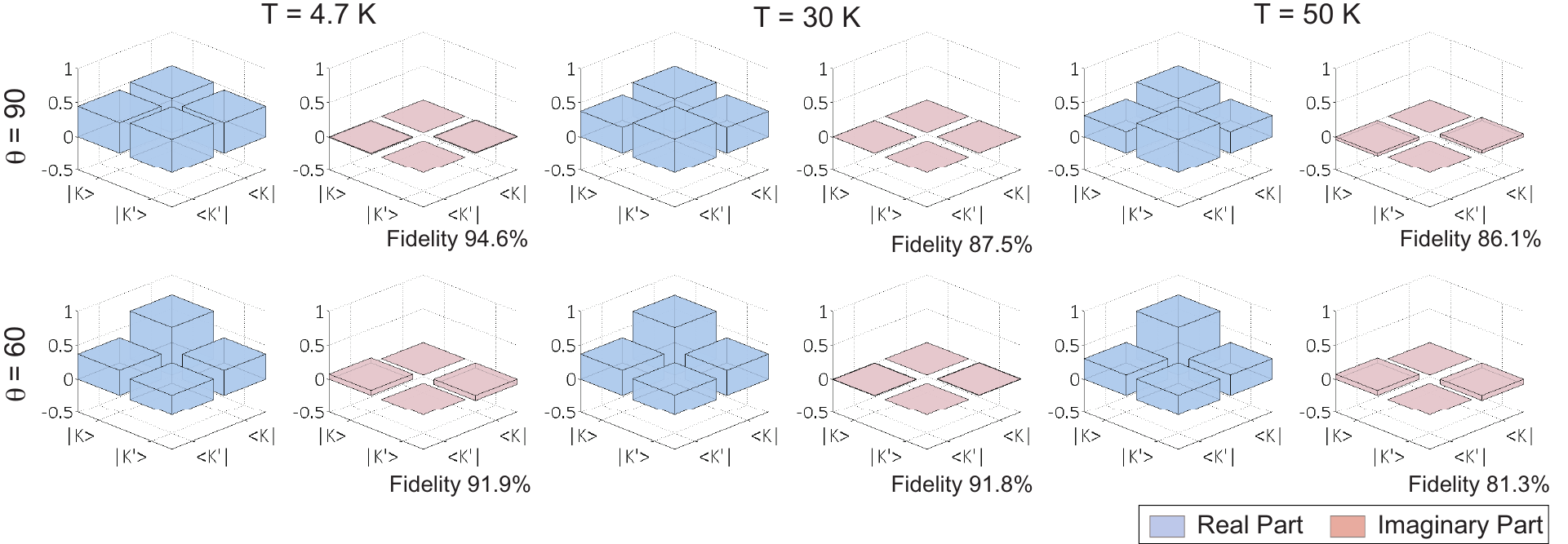}\\
\caption{\textbf{Valley state tomography.} The state tomography results of valley pseudospin superposition
state for different temperatures $T=4.7~K,30~K,50~K$ given two polar angles ($\theta=60^\circ,90^\circ$)
and zero azimuthal angle. The fidelity is defined as $F(\rho_1,\rho_2)=\textrm{Tr}(\sqrt{\sqrt{\rho_1}\rho_2\sqrt{\rho_1}})$ for any two states
$\rho_1$ and $\rho_2$}\label{fig:4}
\end{figure*}

Preparation and a complete readout of a qubit state by measurement are necessary for quantum information processing. In general, the readout can be done by using state tomography to obtain all information of
the state density matrix. Besides the operators in the poles of Bloch sphere as already been performed in
literatures for complete state tomography, we need also the measurement with vectors in the equator to detect the valley quantum coherence and the angle dependent linear polarization PL intensities can be recorded.
The states in the equator of the Bloch sphere are generally named as equatorial qubits in quantum information.
This measurement corresponds to a series of projectors, $\hat{\Pi}_{\alpha}\equiv|\psi_{\frac{\pi}{2},{2\alpha}}\rangle\langle\psi_{\frac{\pi}{2},2\alpha}|$, $|\psi_{\frac{\pi}{2},2\alpha}\rangle=(|K\rangle+e^{i2\alpha}|K'\rangle)/\sqrt{2}$
 with detection angle $\alpha$ ranging from $0^\circ$ to $360^\circ$,
see Fig.~\ref{fig:2}(a). The z-axis rotation of the valley qubit in the Bloch sphere with an angle $2\alpha$ will cause the PL polarization to rotate with {only} 
the angle $\alpha$ \cite{Wang2016,Ye2016}. Given different valley states, we plot the PL intensity
at different temperatures ($T=4.7$~K, 30~K, 50~K) as a function of detection angles $\alpha$ in Fig.~\ref{fig:2}(b-d).
 For the projective measurement operator $\hat{\Pi}_\alpha$ on any pure state $|\psi_{\theta,\phi}\rangle$, the measured
result could be
$p(\alpha)
=[1+\sin\theta\cos(\varphi-2\alpha)]/2$. 
By normalizing the neutral exciton PL intensity, we can obtain the populations for a series of projective measurements,
see Supplementary Information for details.

Then, we can calculate the state tomography results of a valley qubit state. The diagonal elements
are obtained by normalized population results of polar projective measurement. The nondiagonal elements
are calculated by using the least square method in analyzing the normalized PL results.
(See Supplementary Information for details.) With two polar angles $\theta= 60^\circ,90^\circ$ and
zero azimuthal angle, the state tomography results of valley pseudospin superposition states for different temperatures are shown in Fig.~{\ref{fig:4}}.

\textbf{Demonstration of uncertainty relations using normalized PL intensity.}
One direct verification of the quantum coherence of valley coherence is to demonstrate the uncertainty relations, which are closely related with no-cloning theorem
of quantum information \cite{Fan2014}. The uncertainty relation that bounds the uncertainties about the outcomes
of two incompatible observables on one particle was firstly introduced by Heisenberg using the
standard deviation \cite{Heisenberg}, which is the widely accepted form called the Heisenberg-Robertson relation \cite{HR}. In an information-theoretic context of quantum information, the
uncertainty principle can be formulated as the entropic uncertainty relation \cite{eu1,eu2}, which can also be regarded as
a quantum coherence \cite{coherence} induced uncertainty relation, see Supplementary Information for details.

\begin{figure}[b]
 \centering
\includegraphics[width=0.45\textwidth]{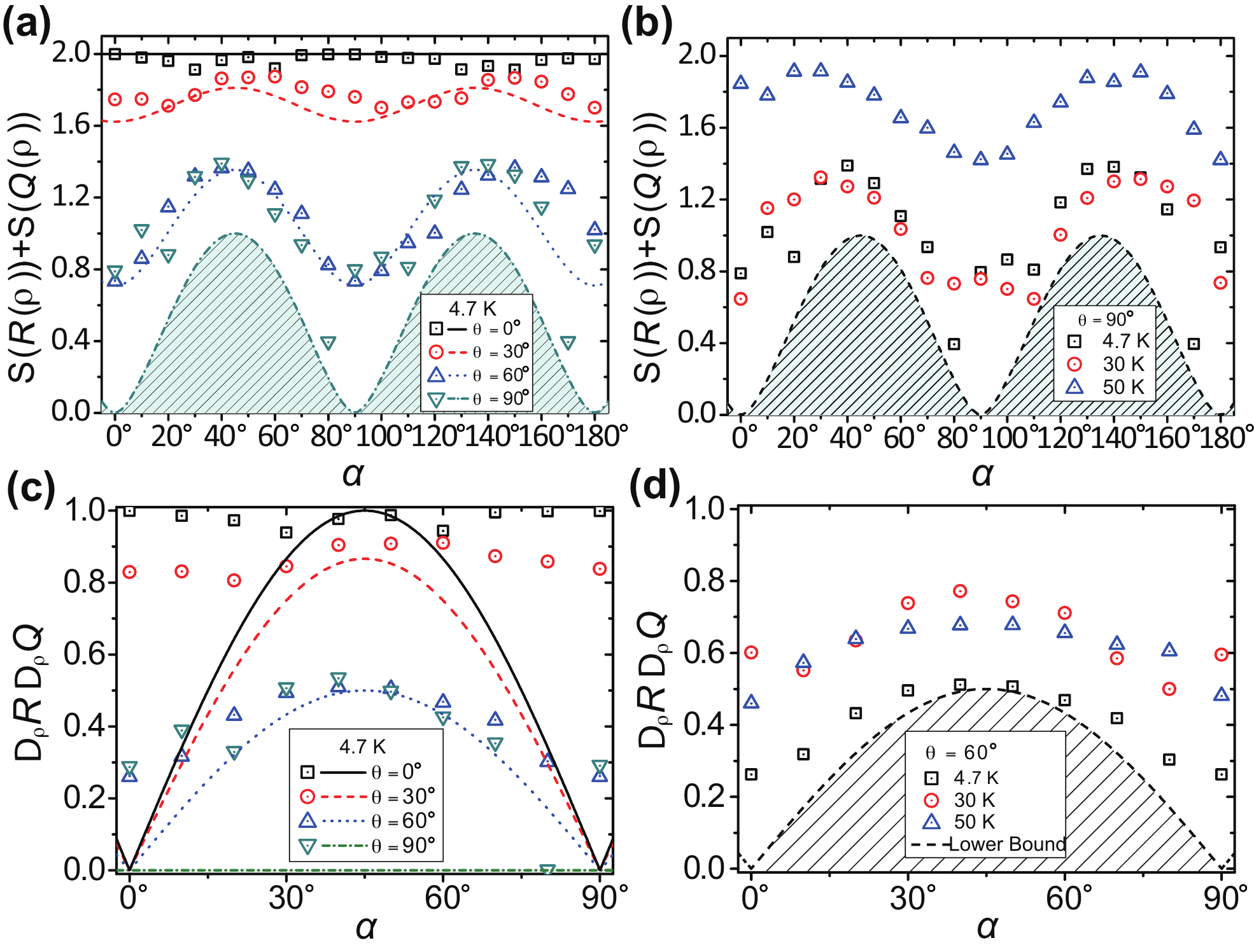}\\
\caption{\textbf{Valley uncertainty relations.}
(a) Entropy uncertainty $S(\hat{R}(\rho))+S(\hat{Q}(\rho))$ against the detection angle $\alpha$ at $T=4.7$~K given four valley states, $\rho=|\psi_{\theta,0}\rangle\langle\psi_{\theta,0}|$ with $\theta=0^\circ,30^\circ,60^\circ,90^\circ$. (b) Given
a valley state $\rho=|\psi_{\frac{\pi}{2},0}\rangle\langle\psi_{\frac{\pi}{2},0}|$,  entropy uncertainty $S(\hat{R}(\rho))+S(\hat{Q}(\rho))$ against
the detection angle $\alpha$ at different temperatures $T=4.7$~K, 30~K and 50~K.
(c) Uncertainty $\Delta_\rho \hat{R}\Delta_\rho \hat{Q}$ against the detection angle $\alpha$ at $T=4.7$~K given four valley states,
$\rho=|\psi_{\theta,0}\rangle\langle\psi_{\theta,0}|$ with $\theta=0^\circ,30^\circ,60^\circ,90^\circ$. (d) Given
a valley state $\rho=|\psi_{\frac{\pi}{3},0}\rangle\langle\psi_{\frac{\pi}{3},0}|$,  uncertainty $\Delta_\rho \hat{R}\Delta_\rho \hat{Q}$
against the detection angle $\alpha$ at different temperatures $T=4.7$~K, 30~K and 50~K. The forbidden shaded area
depends on the lower bound $|\langle[\hat{R},\hat{Q}]\rangle_\rho|/2$.}\label{fig:5}
\end{figure}

Here, given different valley states, we choose two observables as $\hat{R}=\hat{\Pi}_0-\hat{\Pi}_0^{\perp}$ and $\hat{Q}=\hat{\Pi}_\alpha-\hat{\Pi}_\alpha^{\perp}$ with detection angle
$\alpha$ and show the entropy uncertainty $S(\hat{R}(\rho))+S(\hat{Q}(\rho))$ (points) against $\alpha$
compared with the lower bounds and theoretic results (lines) in Fig.~\ref{fig:5}(a,b), where the forbidden shaded area depends on
the lower bound $\log_2[{1}/{c(\hat{\Pi}_0,\hat{\Pi}_\alpha)}]$.
Then, for completeness, we study the Heisenberg uncertainty relation  in ML WSe${}_2$. As $\hat{R}^2=\hat{Q}^2=\mathbb{I}$,
we obtain that $\Delta_\rho \hat{R}^2\Delta_\rho \hat{Q}^2={(1-\langle \hat{R}\rangle^2)(1-\langle \hat{Q}\rangle^2)}$, where the
average reads $\langle \hat{Q}\rangle_\rho\equiv\textrm{Tr}(\hat{Q}\rho)=2p(\alpha)-1$ with probability of projector
$p(\alpha)=\textrm{Tr}(\hat{\Pi}_\alpha\rho)$, and the derivation is defined as
$\Delta_\rho \hat{R}\equiv(\Delta_\rho \hat{R}^2)^{\frac{1}{2}}$ with variance $\Delta_\rho \hat{R}^2\equiv{\langle \hat{R}^2\rangle_\rho-\langle \hat{R}\rangle_\rho^2}$.
In Fig.~\ref{fig:5}(c,d), we plot the Heisenberg uncertainty relation of valley superposition states (points) at
different temperatures against the detection angle $\alpha$ in comparison with the theoretic lower bounds (lines),
see Supplementary Information for details.



\textbf{Summary and discussion.}
In this work, we demonstrate an arbitrary qubit preparation by using the specific
elliptical light excitation with combination of the corresponding left- and right-circularly
lights. In addition, we achieve the complete readout of the qubit by the state tomography by
measuring the circular polarization and intensities of the polarization-resolved PL with
respect to different exciton angles. Thus, with this standard tomography technique, the density
matrix of the valley qubit can be measured quantitatively.
To confirm the quantum coherence of the valley qubit, the unique characteristic
of quantum mechanics, we show that the valley qubit can demonstrate the Heisenberg
uncertainty principle in two different forms: the Heisenberg-Robertson relation and the entropic uncertainty relation.
Our results may pave the way for the valley pseudospin as a qubit which is the
fundamental element acting as the carrier of quantum information. With developments
on coherent manipulation of valley pseudospin states \cite{Wang2015,Wang2016,Schmidt2016,Arora2016,Kim2014,Sie2015,Aivazian2015,Srivastava2015,MacNeil2015},
 more researches by valley qubit can be expected in studying fundamentals of quantum physics and applications in quantum
 computation and quantum information processing.


\textbf{Acknowledgements.}
We would like to thank Xavier Marie and Gang Wang for providing the data for valley dynamics presented in the 
Supplementary Information.
This work was supported by the Ministry of Science and Technology of China (No. 2016YFA0302104, No. 2016YFA0300601, No. 2016YFA0300604, No. 2015CB921001), the National Natural Science Foundation of China ({No. 11574357}, and No. 91536108), and the Chinese Academy of Sciences (No. XDB01010000, and No. XDB21030300).

\textbf{Authors contributions.}
B.L. and H.F. designed the experiment. B.L. is in charge of the experiment,
H.F. is in charge of the theory. J.W., {C.Z.} and B.L. performed the experiment. Y.-R.Z.
and H.F. carried out the theoretical study. Y.-R. Z. wrote the paper with assistance of
J.W., {C.Z.}, B.L. and H.F. The first two authors J.W. and Y.-R.Z. contributed equally to this paper.
All authors analysed the data and commented on the manuscript.
\subsection{Additional information}
\textbf{Competing financial interests:} The authors declare no competing financial interests.

\renewcommand{\theequation}{S\arabic{equation}}
\setcounter{equation}{0}
\renewcommand{\thefigure}{S\arabic{figure}}
\setcounter{equation}{0}
\renewcommand{\thetable}{S\arabic{table}}
\setcounter{equation}{0}

\clearpage

\section{Supplementary Information}

\subsection{Normalized projective measurement on the equator of the Bloch sphere}
\begin{figure}[b]
 \centering
\includegraphics[width=0.4\textwidth]{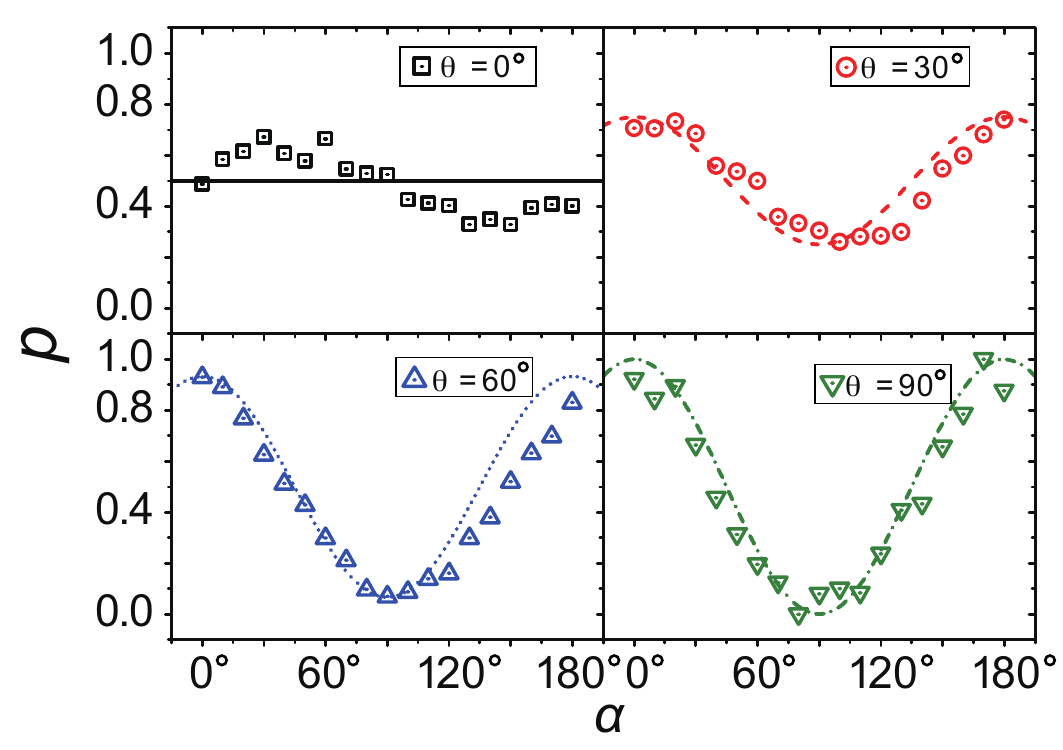}\\
\caption{\textbf{Normalized readout of the valley equatorial detection probabilities.}
Experimental results of  equatorial detection probabilities for elliptically polarized exciton angles
$\theta=0^\circ$, $30^\circ$, $60^\circ$, $90^\circ$ at 4.7~K, compared with theoretic predictions.
}\label{fig:6}
\end{figure}

We denote $|{G}\rangle$ to  the common ground state for excitons in both $K$ and $K'$ valleys. 
The creation and emission of the valley excitons obey the optical selection rules \cite{Ye2016}
$\langle {G}| \bm{\mathscr{P}}\cdot\hat{\sigma}^{+}|K\rangle=\langle {G}| \bm{\mathscr{P}}\cdot\hat{\sigma}^{-}|K'\rangle=\mathscr{D}$
and $\langle {G}| \bm{\mathscr{P}}\cdot\hat{\sigma}^{-}|K\rangle=\langle {G}| \bm{\mathscr{P}}\cdot\hat{\sigma}^{+}|K'\rangle=0$
where $\bm{\mathscr{P}}$ is the electric dipole operator, $\hat{\sigma}^{\pm}=(\hat{\sigma}^X\pm i\hat{\sigma}^Y)/\sqrt{2}$
is the circular polarization direction of the electric field, and $\mathscr{D}$ is the magnitude of the
transition dipole moment for both valleys. For any pure state of valley state
\begin{eqnarray}
|\psi_{\theta,\varphi}\rangle=\cos\frac{\theta}{2}|K\rangle+\sin\frac{\theta}{2}e^{i\varphi}|K'\rangle.\label{eq:1}
\end{eqnarray}
and degenerate excitons in $K$ and $K'$ valleys without control, the total
fluorescence intensity with linearly polarized vector $\hat{\sigma}^\alpha=\cos\alpha\hat{\sigma}^X-\sin\alpha\hat{\sigma}^Y$
may be written as
\begin{eqnarray}
I_\theta(\alpha)\propto|\langle {G}|\bm{\mathscr{P}}\cdot\hat{\sigma}^\alpha|\psi\rangle|^2=\frac{|\mathcal{D}|^2}{2}[1+\sin\theta\cos(\varphi-2\alpha)]
\end{eqnarray}
With two general decay mechanisms, we can obtain the collective fluorescence intensity as
\begin{eqnarray}
I_\theta(\alpha)&\propto&\int_0^{\infty}\!\!\!dt\;|\mathcal{D}|^2[e^{-t/T_1}+e^{-t/T_2^*}\sin\theta\cos(\varphi-2\alpha)]/2\nonumber\\
&\propto&[1+(T_2^*/T_1)\sin\theta\cos(\varphi-2\alpha)]/2
\end{eqnarray}
where $T_1$ is the exciton population lifetime, $T_2^*\equiv1/(1/T_1+1/T_2)$ with $T_2$ the valley exciton coherence time, and the ratio may be calculated as $T_2^*/T_1\simeq0.2$ at 4.7~K.
Technically, we can normalize the PL intensity and obtain
the probability distribution for the measurement
$\hat{\Pi}_\alpha\equiv|\psi_{\frac{\pi}{2},\alpha}\rangle\langle\psi_{\frac{\pi}{2},\alpha}|$ on state (\ref{eq:1}) as
\begin{eqnarray}
p_\theta(\alpha)\equiv(r{I}_{\theta}(\alpha)-I^{\textrm{min}}_{\frac{\pi}{2}})/(I^{\textrm{max}}_{\frac{\pi}{2}}-I^{\textrm{min}}_{\frac{\pi}{2}})
\end{eqnarray}
where $r\equiv(I^{\textrm{max}}_{\frac{\pi}{2}}+I^{\textrm{min}}_{\frac{\pi}{2}})/(I^{\textrm{max}}_{\theta}+I^{\textrm{min}}_{\theta})$, $I_\theta^{\textrm{min}}$ and $I_\theta^{\textrm{max}}$ denote to the minimal and maximal intensities
of PL  for different detection angles. Compared with the ideal theoretic prediction
\begin{eqnarray}
p(\alpha)
=[1+\sin\theta\cos(\varphi-2\alpha)]/2,\label{eq:4}
\end{eqnarray}
the experimental results of the equatorial detection probabilities are shown in
Fig.~\ref{fig:6} for elliptically polarized exciton angles $\theta=0^\circ$, $30^\circ$, $60^\circ$, $90^\circ$
at 4.7~K.

\subsection{Valley state tomography by normalized PL intensity and circular polarization degree.}
For the elliptically polarized exciton light corresponding to the angle $\theta$, the intensity
of light detected in the left-circularly and right-circularly polarized vectors may be divided to three parts
$I=I_\textrm{th}+I_\textrm{pol}+I_\textrm{PL}$,
where  $I_\textrm{th}={I_1}({}^{1}_{0}\ {}^0_1)$
denotes to the coherency matrix of the unpolarized thermal light, $I_\textrm{pol}={I_2}({}^{1}_{1}\ {}^1_1)$ stands for the
polarized thermal light and
\begin{eqnarray}
I_\textrm{PL}={I_3}\left(\begin{array}{c c}1+\cos\theta&e^{-\Gamma-i\varphi}\sin\theta \\ e^{-\Gamma+i\varphi}\sin\theta &1-\cos\theta\end{array}\right)
\end{eqnarray}
corresponds to PL where the valley qubit state may be a mixed state and the temperature decay with a rate $\Gamma>0$ is assumed. The intensity of detected light may be described as
$I(\sigma^\pm)=I_1+I_2+(1\pm\cos\theta)I_3$. The circular PL polarization may be calucalated as
$\eta_\textrm{C}=q_3\cos\theta$,
where the proportion of three different parts of light is defined as $q_i\equiv {I_i}/(I_1+I_2+I_3)$ for $i=1,2,3$.
As $\eta_\textrm{C}\propto\cos\theta$ and using the linear-regression analysis method,
the diagonal elements of the density matrix of the valley state
are retrieved using the results of circular polarizations of PL for different exciton angles in TABLE~\ref{tab:11}.

Given the measurement on the equator of the Bloch sphere
$\hat{\Pi}_\alpha\equiv|\psi_{\frac{\pi}{2},2\alpha}\rangle\langle\psi_{\frac{\pi}{2},2\alpha}|$,
the detected PL may be written as
\begin{eqnarray}
I_{\alpha}(\theta)=I_1+I_3[1+\sin\theta\cos(\varphi-2\alpha)e^{-\Gamma}]
\end{eqnarray}
where we have neglected the polarized part of the thermal light ($I_2\simeq0$).
As discussed in the previous section, we can normalize the PL detected intensity and obtain
the normalized probability.
One general form of the the probability of equatorial detection may be written as
\begin{eqnarray}
p=\frac{1}{2}+\frac{1}{2}\sin\theta e^{-\Gamma}[\cos(2\alpha)\cos\varphi+\sin(2\alpha)\sin\varphi]
\end{eqnarray}
Then, to retrieve the nondiagonal elements of the density matrix of the valley state, we use the
least-square method to calculate the real part $\sin\theta\cos\varphi e^{-\Gamma}/2$ and imaginary
part $\sin\theta\sin\varphi e^{-\Gamma}/2$ given detection angles $\alpha=0^\circ\sim180^\circ$.

\subsection{Demonstration of valley qubit dynamics}
\begin{figure}[b]
 \centering
\includegraphics[width=0.43\textwidth]{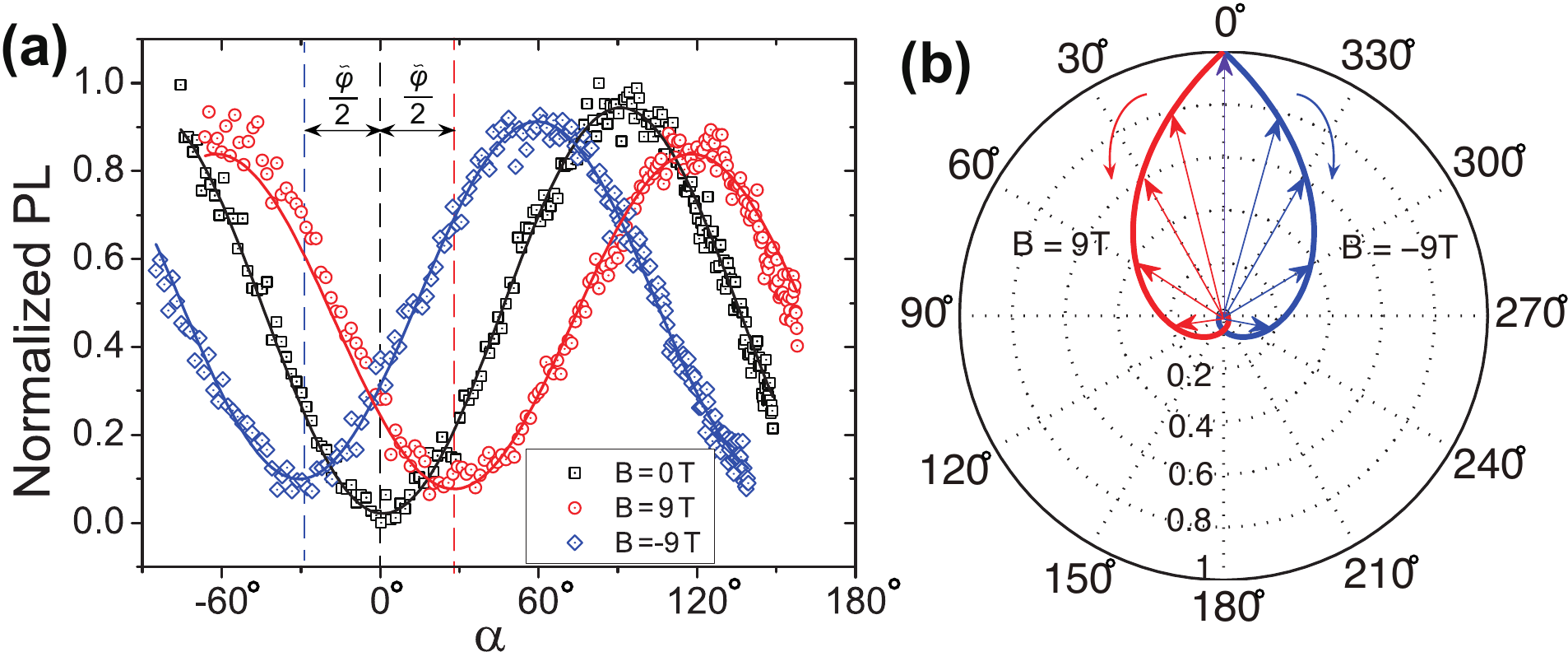}\\
\caption{\textbf{Dynamics of valley qubit by a longitudinal magnetic field.}
(a) The normalized angle dependent PL intensity of the valley exciton for $B=0$~T (black square), $B=9T$ (red circle) and $B=-9$~T (blue triangle).  (b) Representation of valley qubit dynamics on the equatorial plane of Bloch sphere
for $B=9T$ (red) and $B=-9$~T (blue).
The valley qubit state vector is rotated with angular velocity $\Omega\simeq2.93\times10^{12}$~rad/s
The decoherence rate can be written as $e^{-t/T_2^*}$ with the effective coherence time $T_2^*\simeq0.37$~ps \cite{Wang2016}.
}\label{fig:7}
\end{figure}

Universal control of valley qubit is essential for its various applications in quantum information processing and quantum computations.
The manipulation of valley pseudospin can be realized using longitudinal magnetic field \cite{Wang2016,Schmidt2016} and optical Stark effect \cite{Ye2016}. The valley neutral exciton in ML WSe${}_2$ can be controlled by an external magnetic field that is vertical to the sample plane \cite{Wang2016}.
For linearly polarized laser excition $\theta=\pi/2$ and a longitudinal magnetic field $B$, the valley peusdospin evolves with time as $(e^{-i\Omega t/2}|K\rangle+e^{i\Omega t/2}|K'\rangle)/\sqrt{2}$ and the two valleys become nondegenerate with energy difference $\Omega=g\mu_BB\hbar^{-1}$ with $\mu_B$ the Bohr  magneton and $g\simeq-3.7$ the Land\'{e} $g$ factor \cite{Wang2015}.
Then, the polarization-resolved PL intensity can be written as
\begin{eqnarray}
I(\alpha)&\propto&\int_0^{\infty}\!\!\!dt\;[e^{-t/T_1}+e^{-t/T^*_2}\cos(\Omega t-2\alpha)]/2\nonumber\\
&\propto&\frac{1}{2}+\frac{T^*_2}{2T_1[1+(\Omega T_2^*)^2]^{1/2}}\cos(\tilde{\varphi}-2\alpha).
\end{eqnarray}
Thus, we can obtain that
the PL intensity under a longitudinal magnetic field is rotated by an angle $\tilde{\varphi}/2$ with $\tilde{\varphi}\equiv\arctan(\Omega T_2^*)$ compared with the stationary PL intensity, see Fig.~\ref{fig:7}(a), from
which the details of valley peusdospin dynamics can be retrieved, see  Fig.~\ref{fig:7}(b).

\subsection{Quantum coherence and uncertainty relations.}
The uncertainty relation that bounds the uncertainties about the outcomes
of two incompatible observables on one particle was firstly introduced by Heisenberg using the
standard deviation \cite{Heisenberg}. One widely accepted form of this relation
is expressed by the Heisenberg-Robertson relation \cite{HR}: $\Delta \hat{R}\Delta \hat{Q}\geq|\langle[\hat{R},\hat{Q}]\rangle|/2$
where $\Delta \hat{R}$ is the standard deviation of an observable $\hat{R}$. As this
form of relations is state-dependent on the right-hand-side, an improvement of uncertainty relations,
in an information-theoretic context, was subsequently proposed and expressed as
\cite{Kraus,Maassen} $H(\hat{R})+H(\hat{Q})\geq\log_2[{1}/{c(\hat{R},\hat{Q})}]$ where $H(\hat{R})$ denote and the Shannon
entropy of the probability distribution of the outcomes when $\hat{R}$ is measured and
$c(\hat{R},\hat{Q})\equiv\max_{j,k}|\langle r_j|q_k\rangle|^2$ given $|r_j\rangle$ and $|q_k\rangle$ the
eigenvectors of $\hat{R}$ and $\hat{Q}$, respectively.

The relative entropy of coherence \cite{coherence} takes the form, $C(\hat{M})\equiv S(\rho_{\textrm{diag}} )-S(\rho)$,
where $\rho$ is the density matrix, $\rho_{\textrm{diag}}$ keeps only the diagonal elements of $\rho$ in a
fixed basis $\hat{M}$, for example the computational basis $\{|j \rangle\}$, $S(\cdot)$ is the von Neumann entropy.
The physical implication of this definition is that the coherence of a quantum state can be interpreted as the
entropy creating in the measurement $\hat{M}$ corresponding to the fixed basis subtracting the original entropy
\cite{coherence}
\begin{eqnarray}
C(\hat{M})\equiv S(\hat{M}(\rho))-S(\rho)
\end{eqnarray}
where $\hat{M}(\rho)=\sum_j|j\rangle\langle j|\rho|j\rangle\langle j|$  is the post-measurement state which is the
diagonal matrix in a fixed basis and equals to the  Shannon entropy $H(\hat{M})$.
For projective measurement operator in the equator of Bloch sphere $\hat{M}=\hat{\Pi}_\alpha$ and initial state
(\ref{eq:1}), the von Neumann entropy of $\hat{M}(\rho)$ can
be written as a binary entropy ${S}(\hat{M}(\rho))=H_b(p)=-p\log p-(1-p)\log(1-p)$ where $p=[1+\sin\theta\cos(\varphi-\alpha)]/2$.
By applying this observation to a set of measurement operators $\{\hat{M}_k\}$,
we may find that total coherence in different measurements bases is larger than a bound, $\sum_kC(\hat{M}_k )\geq B(\{\hat{M}_k\},\rho)$.
For example, using the entropy uncertainty, we can obtain a lower bound of coherence in
different basis as
\begin{equation}
C(\hat{R})+C(\hat{Q})\geq\log_2[{1}/{c(\hat{R},\hat{Q})}]+2S(\rho)
\geq\log_2[{1}/{c(\hat{R},\hat{Q})}]
\end{equation}
where the first equal condition is the same as the one of entropic uncertainty and the second equality is saturated for pure states.

\begin{table}[b]
\begin{center}
\begin{tabular}{c| cc c c c}
\hline
\hline
 &~Polarization~&$\theta=0^\circ$&$\theta=30^\circ$&$\theta=60^\circ$&$\theta=90^\circ$\\
\hline
~Excitation~&$\eta_\textrm{L}$&14.3\%	&50\%	&86.6\%	&99.6\%\\
laser&$\eta_\textrm{C}$&99\%	&86.6\%	&50\%	&8.7\%\\
\hline
\multirow{2}{*}{PL}
&$\eta_\textrm{L}$&14.3\%&	20\%&	36\%&	41.8\%\\
&$\eta_\textrm{C}$&37.3\%&	35.4\%&	16.3\%&	0\%\\
\hline
\hline
\end{tabular}
\end{center}
\caption{\label{tab:11} {Experimental results of degrees of both circular polarization and linear polarization of linear exciton laser and PL.}}
\end{table}



\begin{thebibliography}{99}
\bibitem{Novoselov2005} Novoselov, K. S. \emph{et al.}
    {Two-dimensional atomic crystals},
     Proc. Natl Acad. Sci. \textbf{102}, 10451-10453 (2005).

\bibitem{Schaibley2016} Schaibley, J. R., Yu, H. Y., Clark, G.,  Rivera,P., Ross, J. S., Seyler, K. L., Yao W. \& Xu, X. D.,
    {Valleytronics in 2D materials},
    Nat. Rev. Mats. \textbf{1}, 16055 (2016).

\bibitem{Rycerz2007} Rycerz, A.,  Tworzyd{\l}o, J., \& Beenakker, C. W.
    {Valley filter and valley valve in graphene},
     Nat. Phys. \textbf{3}, 172-175 (2007).

\bibitem{Shkolnikov2002} Shkolnikov, Y., De Poortere, E., Tutuc, E. \& Shayegan, M.
    {Valley splitting of AlAs two-dimensional electrons in a perpendicular magnetic field},
    Phys. Rev. Lett. \textbf{89}, 226805 (2002).

\bibitem{Gunawan2006} Gunawan, O. \emph{et al.}
    {Valley susceptibility of an interacting two-dimensional electron system},
    Phys. Rev. Lett. \textbf{97}, 186404 (2006).

\bibitem{Cao2012} Cao, T. \emph{et al.}
    {Valley-selective circular dichroism of monolayer molybdenum disulphide},
    Nat. Commun. \textbf{3}, 887 (2012).

\bibitem{Mak2012} Mak, K. F., He, K., Shan, J. \& Heinz, T. F.
    {Control of valley polarization in monolayer MoS${}_2$ by optical helicity},
     Nat. Nanotechnol. \textbf{7}, 494-498 (2012).

\bibitem{Zeng2012} Zeng, H., Dai, J., Yao, W., Xiao, D. \& Cui, X.
    {Valley polarization in MoS${}_2$ monolayers by optical pumping},
     Nat. Nanotechnol. \textbf{7}, 490-493 (2012).

\bibitem{Jones2013} Jones, A. M. \emph{et al.}
    {Optical generation of excitonic valley coherence in monolayer WSe${}_2$},
    Nat. Nanotechnol. \textbf{8} 634-638 (2013)

\bibitem{Xiao2012} Xiao, D., Liu, G.-B., Feng, W., Xu, X. \& Yao, W.
    {Coupled spin and valley physics in monolayers of MoS${}_2$ and other group-VI dichalcogenides},
     Phys. Rev. Lett. \textbf{108}, 196802 (2012).

\bibitem{Zhu2012} Zhu, Z., Collaudin, A., Fauque, B., Kang, W. \& Behnia, K.
    {Field-induced polarization of Dirac valleys in bismuth},
    Nature Phys. \textbf{8}, 89-94 (2012).

\bibitem{Bishop2007} Bishop, N. C. \emph{et al.}
    {Valley polarization and susceptibility of composite fermions around a filling factor $\nu=3/2$},
    Phys. Rev. Lett. \textbf{98}, 266404 (2007).

\bibitem{Chernikov2014} Chernikov, A., Berkelbach, T. C., Hill, H. M., Rigosi, A., Li, T. L., Aslan, O. B., Reichman, D. R., Hybertsen, M. S. \& Heinz, T. F.
    Exciton binding energy and nonhydrogenic Rydberg series in monolayer WS${}_2$,
    Phys. Rev. Lett. \textbf{113}, 076802 (2014).

\bibitem{Ugeda2014} Ugeda, M. M., Bradley, A. J., Shi, S.-F., da Jornada, F. H., Zhang, Y., Qiu, D. Y., Ruan, W., Mo, S.-K., Hussain, Z., Shen, Z. X., Wang, F., Louie, S. G. \& Crommie, M. F.
    Giant bandgap renormalization and excitonic effects in a monolayer transition metal dichalcogenide semiconductor,
    Nat. Mater. \textbf{13}, 1091-1095 (2014).

\bibitem{He2014} He, K. L., Kumar, N., Zhao, L., Wang, Z. F., Mak, K. F., Zhao, H. \& Shan, J.
    Tightly Bound Excitons in Monolayer WSe${}_2$,
    Phys. Rev. Lett. \textbf{113}, 026803 (2014).

\bibitem{Ye2014} Ye, Z. L., Cao, T., O'Brien, K., Zhu, H. Y., Yin, X. B., Wang, Y., Louie, S. G. \& Zhang, X.
    Probing excitonic dark states in single-layer tungsten disulphide,
    Nature \textbf{513}, 214-218 (2014).

\bibitem{Schmidt2016} Schmidt, R., Arora, A., Plechinger, G., Nagler, P., del \'{A}guila, A. G., Ballottin, M. V., Christianen, P. C. M., de Vasconcellos, S. M., Sch\"{u}ller,  C., Korn, T. \& Bratschitsch, R.
    Magnetic-field-induced rotation of polarized light emission from monolayer WS$_{2}$,
    Phys. Rev. Lett. \textbf{117}, 077402 (2016).

\bibitem{Wang2016} Wang, G., Marie, X., Liu, B. L., Amand, T., Robert, C., Cadiz, F., Renucci, P. \& Urbaszek, B.
	Control of exciton valley coherence in transition metal dichalcogenide monolayers,
	Phys. Rev. Lett. \textbf{117}, 187401 (2016).

\bibitem{Ye2016} Ye, Z. L., Sun, D. Z. \& Heinz, T. F.
	Optical manipulation of valley pseudospin,
	Nat. Phys. \textbf{13}, 26-29 (2017).

\bibitem{Fan2014} Fan, H., Wang, Y. N., Jing, L., Yue, J. D., Shi, H. D., Zhang, Y. L. \& Mu, L. Z.
	Quantum cloning machines and the applications,
	Phys. Rep. \textbf{544}, 241-322 (2014).

\bibitem{Heisenberg} Heisenberg, W.
    \"{U}ber den anschaulichen Inhalt der quantentheoretischen Kinematik und Mechanik,
    {Zeitschrift f\"{u}r Physik} \textbf{43}, 172 (1927).

\bibitem{HR} Robertson, H. P.
    The uncertainty pricinple,
    {Phys. Rev.} \textbf{34}, 163 (1929).

\bibitem{eu1} Kraus, K. Complementary observables and uncertainty relations. Phys. Rev. D \textbf{35}, 3070-3075
(1987).

\bibitem{eu2} Maassen, H. \& Uffink, J. B. Generized entropic uncertainty relations. Phys. Rev. Lett. \textbf{60},
1103-1106 (1988).

\bibitem{coherence} Baumgratz, T., Cramer, M. \& Plenio, M. B.
    Quantifying coherence,
    Phys. Rev. Lett. \textbf{113}, 140401 (2014).

\bibitem{Wang2015} Wang, G., Bouet, L., Glazov, M. M., Amand, T., Ivchenko, E. L., Palleau, E., Marie, X. \& Urbaszek, B.
    Magneto-optics in transition metal diselenide monolayers,
    2D Mater. \textbf{2}, 034002 (2015).


\bibitem{Arora2016} Arora, A., Schmidt, R., Schneider, R., Molas, M. R., Breslavetz, I., Potemski, M. \& Bratschitsch, R.
	Valley Zeeman splitting and valley polarization of neutral and charged excitons in monolayer MoTe${}_2$ at high magnetic fields,
	Nano Lett. \textbf{16}, 3624-3629 (2016).
	
\bibitem{Aivazian2015} Aivazian, G. \emph{et al.}
	Magnetic control of valley pseudospin in monolayer WSe${}_2$,
	Nat. Phys. \textbf{11}, 148-152 (2015).

\bibitem{Srivastava2015} Srivastava, A., Sidler, M., Allain, A. V., Lembke, D. S., Kis, A. \& Imamo\v{g}lu, A.
	Valley Zeeman effect in elementary optical excitations of monolayer WSe${}_2$,
	Nat. Phys. \textbf{11}, 141-147 (2015)

\bibitem{MacNeil2015} MacNeill, D., Heikes, C., Mak, K. F., Anderson, Z., Korm\'{a}nyos, A., Z\'{o}lyomi, V., Park, J. \& Ralph, D. C.
	Breaking of valley degeneracy by magnetic field in monolayer MoSe${}_2$,
	Phys. Rev. Lett. \textbf{114}, 037401 (2015).

\bibitem{Kim2014} Kim, J., Hong, X., Jin, C., Shi, S.-F., Chang, C.-Y. S., Chiu, M.-H., Li, L.-J. \& Wang, F.
	Ultrafast generation of pseudo-magnetic field for valley excitons in WSe${}_2$ monolayers,
	Science \textbf{346}, 1205-1208 (2014).

\bibitem{Sie2015} Sie, E. J., McIver, J. W., Lee, Y.-H., Fu, L., Kong, J. \&  Gedik, N.
	Valley-selective optical Stark effect in monolayer WS${}_2$,
	Nature Mater. \textbf{14}, 290-294 (2015).





	

\end{thebibliography}

\begin{thebibliography}{99}
\bibitem{Ye2016} Ye, Z. L., Sun, D. Z. \& Heinz, T. F.
	Optical manipulation of valley pseudospin,
	Nat. Phys. \textbf{13}, 26-29 (2017).
	
\bibitem{Wang2016} Wang, G., Marie, X., Liu, B. L., Amand, T., Robert, C., Cadiz, F., Renucci, P. \& Urbaszek, B.
	Control of exciton valley coherence in transition metal dichalcogenide monolayers,
	Phys. Rev. Lett. \textbf{117}, 187401 (2016).

\bibitem{Schmidt2016} Schmidt, R., Arora, A., Plechinger, G., Nagler, P., del \'{A}guila, A. G., Ballottin, M. V., Christianen, P. C. M., de Vasconcellos, S. M., Sch\"{u}ller,  C., Korn, T. \& Bratschitsch, R.
    Magnetic-field-induced rotation of polarized light emission from monolayer WS$_{2}$,
    Phys. Rev. Lett. \textbf{117}, 077402 (2016).

\bibitem{Wang2015} Wang, G., Bouet, L., Glazov, M. M., Amand, T., Ivchenko, E. L., Palleau, E., Marie, X. \& Urbaszek, B.
    Magneto-optics in transition metal diselenide monolayers,
    2D Mater. \textbf{2}, 034002 (2015).

\bibitem{Heisenberg} Heisenberg, W.
    \"{U}ber den anschaulichen Inhalt der quantentheoretischen Kinematik und Mechanik,
    {Zeitschrift f\"{u}r Physik} \textbf{43}, 172 (1927).

\bibitem{HR} Robertson, H. P.
    The uncertainty pricinple,
    {Phys. Rev.} \textbf{34}, 163 (1929).

\bibitem{Kraus} Kraus, K.
    Complementary observables and uncertainty relations,
    {Phys. Rev. D} \textbf{35}, 3070 (1987).

\bibitem{Maassen} Maassen, H. \&  Uffink, J. B.
    Generized entropic uncertainty relations,
    {Phys. Rev. Lett.} \textbf{60}, 1103 (1988).

\bibitem{coherence} Baumgratz, T., Cramer, M. \& Plenio, M. B.
    Quantifying coherence,
    Phys. Rev. Lett. \textbf{113}, 140401 (2014).
	





\end{thebibliography}
\end{document}